# Free-electron superfluorescence: collective optical dynamics at deep-subwavelength resolution

Orr Be'er[1†], Alexey Gorlach[2†], Alina Nagel[1], Reut Shechter[1], Yaniv Kurman[2], Marc Fouchier[3], Ido Kaminer[2], Yehonadav Bekenstein[1*]

[1]Department of Materials Science and Engineering and the Solid-State Institute, Technion − Israel Institute of Technology, 32000 Haifa, Israel

[2]Department of Electrical Engineering and the Solid-State Institute, Technion – Israel Institute of Technology, Haifa, Israel.

[3]Attolight AG / EPFL Innovation Park, Building D, 1015 Lausanne, Switzerland

[†] *These authors contributed equally to this work*

*Corresponding author. Email: bekenstein@technion.ac.il

**Long-range coherence and correlations between electrons in solids are the cornerstones for developing future quantum materials and devices[1,2]. In 1954, Dicke described correlated spontaneous emission from closely packed quantum emitters[3], forming the theoretical basis of superradiance and superfluorescence. Since then, it has remained an open challenge to observe such phenomena with nanometer spatial resolution, precisely the important scale at which the collective correlations occur[4]. Here, we report the first instance of free-electron-driven superfluorescence –** *superfluorescent cathodoluminescence* **– enabling us to excite and observe correlations at nanometer spatial scales. To exemplify this concept in our experiments, superlattices of lead halide perovskite quantum dots[5] are excited by focused pulses of multiple free electrons. The electrons trigger superfluorescence: collective ultrafast emission observed at rates faster than both the lifetime and decoherence time. By controlling the area illuminated by the electron beam, we create a transition from a non-correlated spontaneous emission to a correlated superfluorescent emission. The observed signatures of superfluorescence are a reduction of the intrinsic emitter lifetime, a narrower linewidth, and a distinct redshift. We develop the theory of superfluorescent cathodoluminescence, which matches the results and highlights the unique features of electron-driven versus light-driven superfluorescence. Our observation and theory introduce a novel way to characterize coherence and correlations in quantum materials with nanometer spatial resolution, a key for future engineering of quantum devices.**

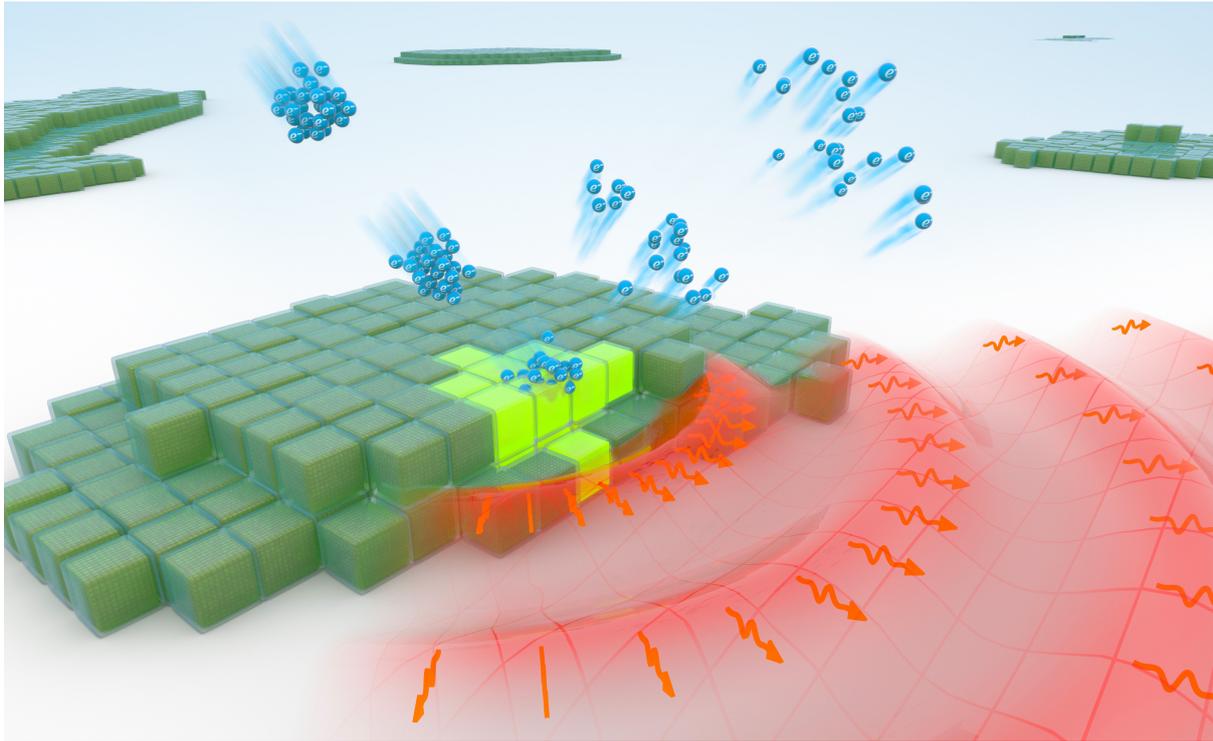

**Superfluorescent cathodoluminescence realized with electron-triggered perovskite quantum dots (QDs).** A pulsed electron beam excites superlattices of QDs. After the interaction, excitonic correlations are gradually built between neighboring QDs. Consequently, the superlattice collectively emits a short pulse of light.

The rapid development of quantum technologies in the optical range is motivating research on novel quantum materials that support quantum-coherent optical dynamics. Particularly important are materials that can sustain long-range coherence and strong correlations between spatially separated emitters. Quantum dots (QDs) are nanometer-size semiconductor crystals that spatially confine their excitons, creating discrete atom-like energy levels and large dipole moments. The dipoles cause strong interactions between excitons and induce correlations between neighboring QDs. Specifically, lead-halide perovskite ($CsPbBr_3$) QDs are attractive quantum light emitters[6] because of their large absorption cross section, efficient emission, and very fast radiative rates[5,7,8]. In closely packed ensembles of such QDs, the large dipoles and correlations of excitons were recently shown to create superfluorescence, observed as a fast coherent burst of radiation following a laser pulse excitation[2].

The theory of correlated collective emission from multiple emitters was first described by Dicke[3]. The collective dynamics can be classified into two distinct regimes[9]: the first is Dicke superradiance, where the initial excitation pulse causes the emitters to become correlated, and the second is superfluorescence, where the correlation between emitters is built up later during the emission process. Both phenomena necessitate long coherence times and a narrow distribution in the emitters' emission spectra. One of the distinct signatures of these effects is an increase in emission rate and overall intensity that scale with the number of correlated emitters[3].

Superfluorescence was extensively studied using light excitation[4,9–13] and demonstrated in several solid-state systems (e.g., Refs.[2,14–18]); however, many of its fundamental properties have remained beyond experimental reach. The challenge arises directly from the stringent conditions required for superfluorescence to occur. These include an orchestrated buildup of correlations between multiple emitters on ultrafast (usually picosecond) time scales, simultaneously with very small (sub-wavelength) distances between the emitters. Observation of superfluorescence phenomena has been limited by the spatial resolution of the experimental probe, bound by the laser wavelength. Notably, the probe laser wavelength is two to three orders of magnitude larger than the typical emitters' sizes and the distances between them, which are the crucial parameters for the buildup of the correlations necessary for superfluorescence.

Free electrons provide a probe with an intrinsic nanometric resolution that can trigger light emission from quantum emitters through a process known as incoherent cathodoluminescence[19–22]. Despite the prevalence of cathodoluminescence techniques, the demonstration of the concept of free-electron-driven superfluorescence has so far remained inaccessible in experiments. Prominent examples of experimentally accessible cathodoluminescence processes include Cherenkov radiation[23], Smith–Purcell radiation[24], and transition radiation[25,26]. These *coherent* cathodoluminescence processes describe radiation emitted from the electrons, mediated by a material with which the electrons interact. Consequently, the radiation has certain coherent properties, such as directionality, a polarized nature, and even a tunable emission profile arising from the spatial shape of the excited electromagnetic mode[27].

In contrast, when free electrons interact with quantum emitters, they trigger another form of radiation known as *incoherent* cathodoluminescence, which depends on the nature of the emitters and their interaction with vacuum fluctuations. The electrons thus provide the energy for the emission rather than emitting the radiation directly. This type of process has never been considered for superfluorescence or shown any type of collective emission, as it is commonly considered to be incoherent.

We present a novel cathodoluminescence effect: superfluorescent cathodoluminescence, in which an ensemble of emitters develops quantum correlations and emits collectively following a free-electron excitation. Despite being a type of incoherent cathodoluminescence, this new form of superfluorescence effect has intrinsic quantum coherence from correlations built up between the emitters.

Our experiment was based on ultrafast cathodoluminescence with time-correlated photon counting, which in recent years has become a powerful tool for mapping material response and emission lifetime on scales that are two orders of magnitude smaller than the emitted wavelength[28–30]. Our electron pulses were a few picoseconds in duration and contained an average of 12 electrons per pulse, focused to nanometer scales inside an ultrafast scanning electron microscope (USEM). We used these electron pulses to excite ensembles of closely packed QDs. The excited QDs interacted and built up correlations on the nanoscale, enabling us to probe the resulting emission properties[31–33].

By controlling the electron pulse excitation, we show how the emission can be switched from uncorrelated spontaneous emission to correlated superfluorescent emission. We develop the theoretical model of superradiant cathodoluminescence, which relies on the fundamental interaction of free electrons with quantum emitters[31], and show that it fits the experimental data. The theory helps identify the threshold between uncorrelated spontaneous and collective superfluorescent emissions, also comparing electron-driven superfluorescence with the conventional light-driven superfluorescence.

**Superfluorescence triggered by free electrons**

To emphasize the importance of deep sub-wavelength resolution for this experiment, we now present a detailed breakdown of the microscope and sample used to realize superfluorescent cathodoluminescence. The ultrafast scanning electron microscope used in the experiment was designed around a high numerical aperture reflective optics used to collect the light and direct it to a streak camera for spectral and temporal analysis (Fig. 1(a)). The sample consists of quantum dot superlattices imaged by an electron microscope (for synthesis and characterization see Supplementary Information), as shown in three magnifications that highlight the proximity of the individual QD emitters, thereby which enable them to develop correlations (Fig. 1(b)). The quantum dot superlattices are held at liquid helium cryogenic temperatures and are excited by ultrafast electron pulses. The resulting superfluorescent emission from the samples is collected and analyzed. To understand the experimental results better, we first describe the theoretical framework and related predictions.

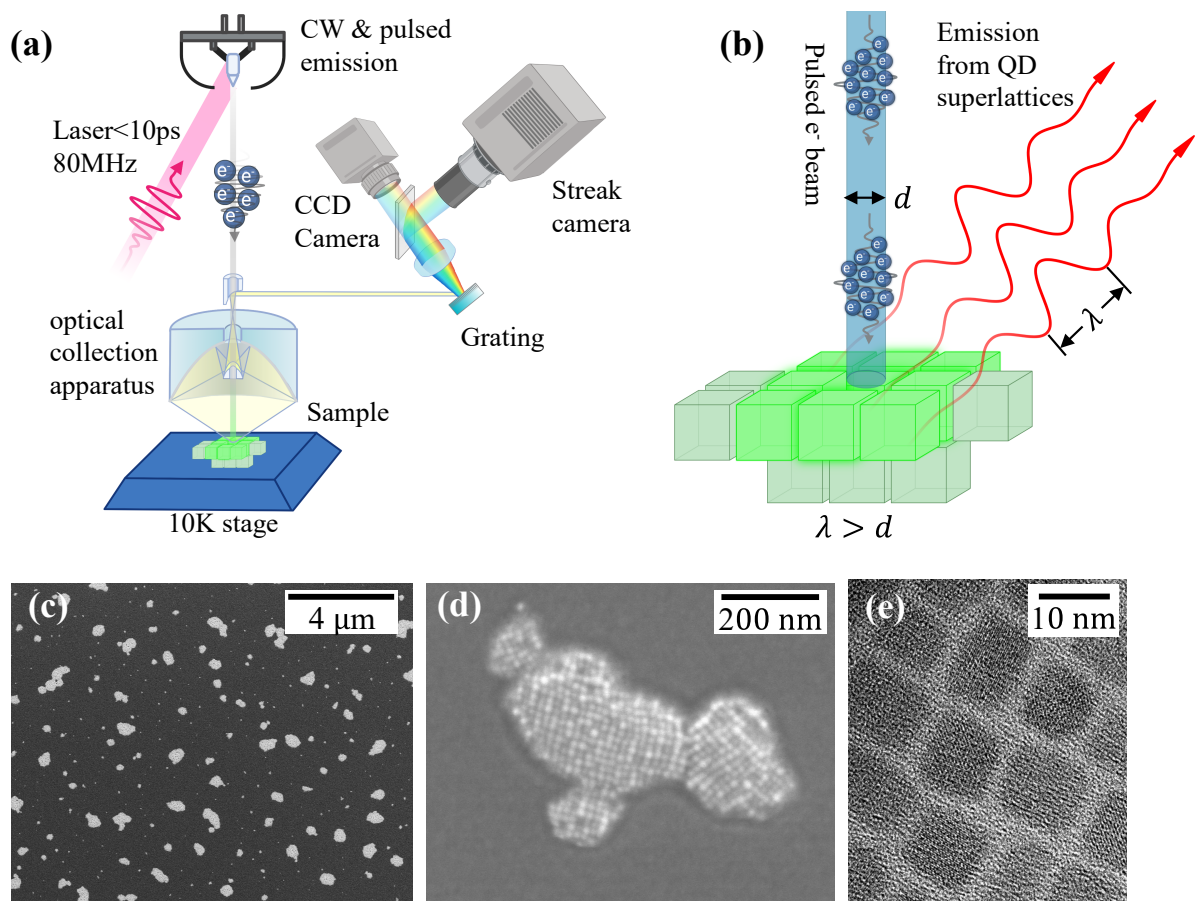

**Fig 1. Superfluorescent cathodoluminescence from a superlattice of quantum dots (QDs): experimental setup. (a)** Ultrafast scanning electron microscope (USEM) for time-resolved cathodoluminescence. Emission from a closely packed spatially ordered superlattice of lead-halide QDs (the sample) is triggered by a pulsed electron beam. **(b)** The electron beam is focused to a beam size $d$ on the nanometer scale, much smaller than the emitted wavelength λ. **(c)** Low magnification image of multiple QD superlattices. The dark background is the polymer (polystyrene protecting matrix), and the white areas are the QDs superlattices. The sizes of the ordered superlattices vary between a few tens of nanometers and a few microns. **(d)** High magnification image of one superlattice, in which separated QDs are clearly observed. **(e)** TEM micrograph of QDs depicting the crystallinity of the QDs and their typical size distribution.

We built a theoretical model that describes the interaction of electrons with multiple QD emitters. This theory captures the dynamics of the excitation and emission process, in agreement with the experimental results (Fig. 2(h)). We later expanded our theory to capture also the effect of the spatial distribution of emitters and the influence of the electron beam size (Fig. 2(a)). We found that it is sufficient to model each QD as a two-level system by matching the experimental data to theory.

Our theoretical model starts with an interaction between a single electron and a single QD. We then increase the number of electrons and QDs (see Ch. S1-3 in Supplementary Information). Based on Dicke's theory, it is typical to assume that the QDs are located very close to each other, so that all the spatial effects can be neglected (see Ch. S4 in Supplementary Information). Each electron can either excite a two-level system (denoted by $\hat{\sigma}_i^+$) by losing energy (denoted by $\hat{b}$) or quench its energy (denoted by $\hat{\sigma}_i^-$) by gaining energy (denoted by $\hat{b}^\dagger$). We show that collective emission (i.e., superfluorescence) is possible in this case (see Ch. S5 in Supplementary Information). The interaction between the superlattice of QDs and a single free electron can be described by the following scattering matrix $S$:

$$S = e^{-i(g\hat{b}\hat{S}^+ + g^*\hat{b}^\dagger\hat{S}^-)}, \qquad (1)$$

where $\hat{S}^\pm = \sum_i \hat{\sigma}_i^\pm$ is the summation over excitation $\hat{\sigma}_i^+$ and deexcitation $\hat{\sigma}_i^-$ operators of all the QDs.

The states of the quantum system of $N$ QDs located within the interaction volume can be described using the symmetric states basis $|l = 0\rangle, \dots, |l = N\rangle$, where each is defined as a superposition of all states of exactly $l$ excited emitters:

$$\begin{cases} |l = N\rangle = |eee\dots ee\rangle, \\ |l = N-1\rangle = N^{-1/2}(|gee\dots ee\rangle + |ege\dots ee\rangle + \dots + |eee\dots eg\rangle), \\ \dots \\ |l = 0\rangle = |ggg\dots gg\rangle. \end{cases} \qquad (2)$$

$|g\rangle$ and $|e\rangle$ are ground and excited states of the two-level systems. Eqs. (1) and (2) can also be used to describe the interaction of $N$ QDs with $N_e$ electrons. In this case, the scattering $S$ from Eq. (1) is applied for each of $N_e$ electrons separately. Eq. (2) shows that the interaction with electrons can excite the system from the ground state to a higher symmetric state. An excitation of symmetric states can lead to superradiant dynamics[3,9] and result in fast superfluorescent emission analogous to optical excitation. We found that this model successfully captures the dynamics of the excitation and emission that we observe (Fig. 2(h) and Ch. 5 in Supplementary Information). However, it does not capture the influence of the electron beam size.

A significant difference between superfluorescence triggered by free electrons and by light is that the free electrons can be strongly focused to the nanometer scale, whereas light focusing is diffraction-limited. Notably, the Dicke model[3] and its conventional generalizations[4,9–13] cannot describe such an excitation. We, therefore, developed a qualitative model that can capture the spatial shape of the electron beam and the distribution of emitters at these length scales (see Ch. 6 in Supplementary Information). Our analysis of the measured emission dynamics showed that the superfluorescence dynamics can be described by a single lifetime $\tau_{SF}$ parameter, which we can derive from a formula analogous to the one describing light-driven superfluorescence[2]:

$$\tau_{SF} = \frac{\tau_{SE}}{1 + \alpha \cdot \sigma_e \cdot F_e} + \Delta t. \qquad (3)$$

$\tau_{SE}$ is the lifetime of spontaneous emission of a single emitter, $\alpha$ describes the strength of correlations between emitters scaling with the density of emitters (units of inverse energy) independently of the method of excitation, $\sigma_e$ is a cross-section parameter for the electron interaction with the QD, and $F_e$ is the excitation density of the electron pulse, defined as the total energy of the multi-electron pulse divided by the beam size. $\Delta t$ is a delay that encompasses time resolution and other imperfections of the measurement.

For comparison, $F_e$ is analogous to $F_{ph}$, the optical excitation density. Similarly, the cross-section parameter for electron excitation $\sigma_e$ is defined in analogy to the cross section for light excitation $\sigma_{ph}$[2]. Both cross-section parameters $\sigma_e$, $\sigma_{ph}$ are limited from above by the size of the QD superlattice and by the superfluorescence correlation distance. However, their limit from below is substantially different, since the electron beam size is orders of magnitude smaller than that of visible light, enabling us to image superfluorescence with a resolution better than the correlation distance. Our theory shows that the superfluorescence lifetime is indeed dependent on $\sigma_e$ and consequently on the electron beam size, in agreement with the results in Fig. 2(c) and (h).

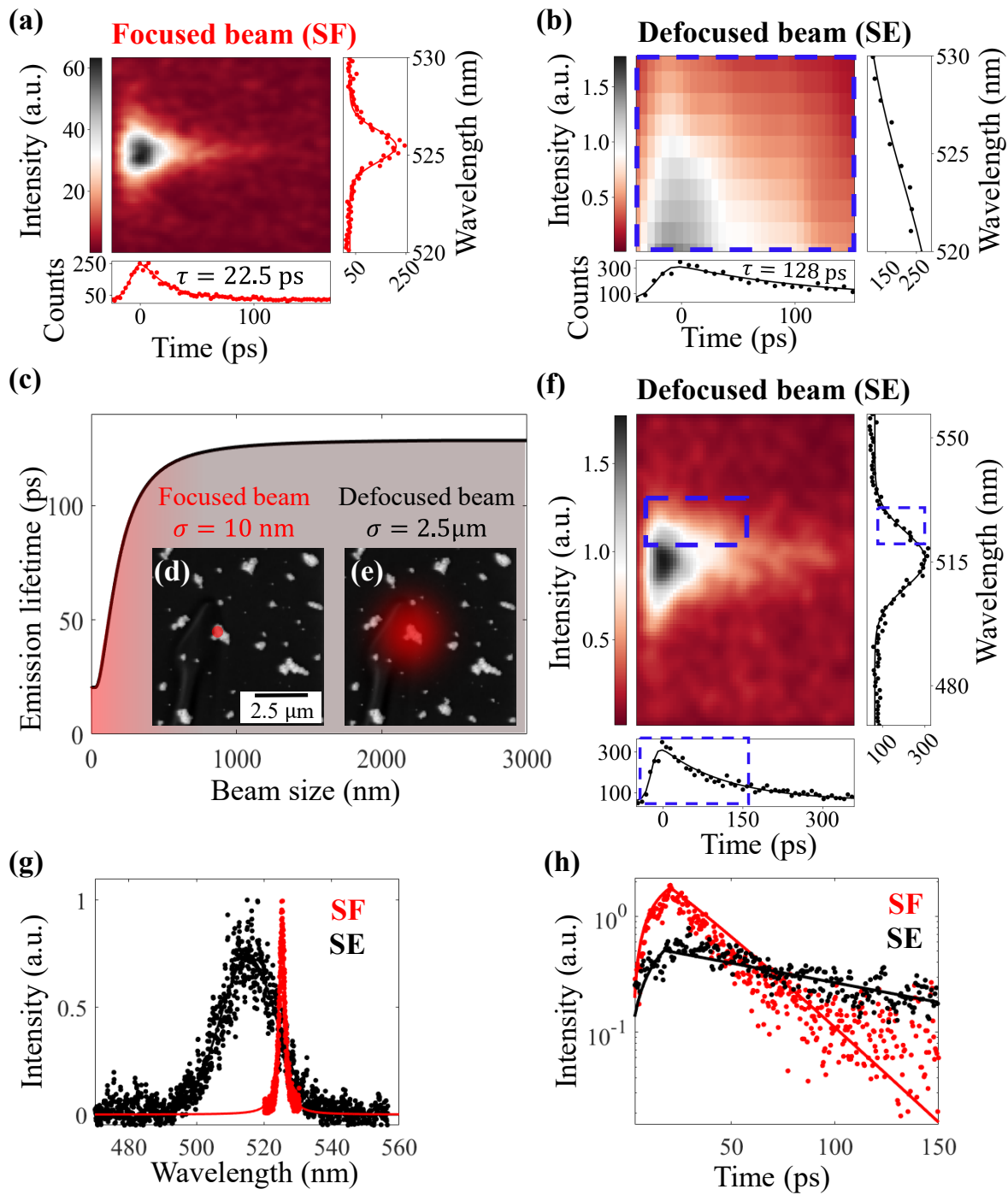

**Fig. 2. Characteristics of electron-driven superfluorescence: emission lifetime and spectrum. (a)** Streak camera image of superfluorescence triggered by a free-electron pulse of 10 nm beam size.

Ultrafast emission lifetime of 22.5 ps and narrow spectral linewidth of 1.9 nm are observed. **(b)** Streak camera image of spontaneous emission triggered by a free-electron pulse of 2.5 μm beam size. The emission lifetime of 128 ps and broader spectral linewidth of 15 nm are observed. **(c)** Theoretically predicted emission time as a function of electron beam size $d$, for 12 electrons per pulse. **(d)–(e)** The beam sizes of $d$=10 nm and $d$=2.5 μm that correspond to our measurements in panels (a) and (b) are denoted on top a low magnification image of the sample. **(f)** Zoom-out of panel (b), showing the streak camera image of spontaneous emission that requires a larger temporal range and spectral range than the case of superfluorescence in panel (a). **(g)** Comparison of spectra from the streak camera images: superfluorescence (red dots) and spontaneous emission (black dots). The superfluorescence spectrum is fitted to a Lorentzian (red curve) and the spontaneous emission spectrum is fitted to a Gaussian (black curve). **(h)** Comparison of a lifetime from the streak camera images: data (dots) and theory (curve) of superfluorescence (red) and of spontaneous emission (black). In both theory and experiment, the emission intensity and rate are enhanced by the number of correlated emitters, as in conventional light-driven superfluorescence.

Our results in Fig. 2 suggest that superfluorescence can emerge beyond the stringent conventional Dicke conditions where all emitters are indistinguishable. We can probe this regime of superfluorescence using an electron excitation that is strongly focused, causing emitters at different distances from the electron beam center to interact differently with the electrons. According to the conventional Dicke conditions, this inhomogeneity in the excitation may be expected to induce multiple emission rates in different areas of the superlattice. However, the overall dynamics (Fig. 2(a) and (h)) fits a single exponential decay rate, which implies that the correlations induce complete synchronization between the different emitters in the superlattice, despite their different distances from the electron beam size. In other words, all the emitters decay simultaneously, and at the same rate, despite experiencing different excitation strengths.

We demonstrated this experimentally by controlling the electron-beam size, comparing two distinct emission regimes: **(1)** The conventional rate of spontaneous emission is measured when the pulsed electron beam is defocused to a larger area (typically micron-scale e.g., $d$=2.5 μm in the case of Fig. 2(e)). For these parameters, we observe emission with a center wavelength peak at $\lambda$=515 nm (Fig. 2(b) and Fig. 2(g) black curve) and a decay time of $\tau_{SE}$=128 ps (Fig. 2(b) and Fig. 2(f) black curve). We attribute this emission to uncoupled and uncorrelated QDs and hypothesize that under these conditions QDs are independently excited and are too distant to interact. **(2)** Superfluorescent emission is measured when the pulsed electron beam is focused to a smaller area on a superlattice (e.g., $d$=10 nm in the case of Fig. 2(d)). For these parameters, we observe a spectrally narrow, redshifted peak at $\lambda$=525 nm (Fig. 2(a) and Fig. 2(g) red curve) and a faster decay time of $\tau_{SF}$=22.5 ps (Fig. 2(a) and Fig. 2(h) red curve). We attribute this emission to superfluorescence from coherently coupled QDs. The transition from uncorrelated spontaneous emission to superfluorescent emission is possible when the electron density is sufficient to excite several QDs separated by distances considerably smaller than the emitted wavelength and sufficient to reach an emission time shorter than the dephasing of the excitons.

We affirm that the emission from the focused beam's excitation is superfluorescent for the four following reasons: **(1)** The emission lifetime is decreased five-fold relative to spontaneous emission. **(2)** The emission spectral linewidth is reduced ten-fold relative to spontaneous emission. **(3)** The emission spectrum fits a Lorentzian line shape, rather than Gaussian line shapes that fit all the measurements of uncorrelated QDs and of low-density excitations. These line shapes agree with the theoretical model of superfluorescence. **(4)** The emission spectrum is redshifted by ~10 nm. Recently, authors have questioned superfluorescence from CsPbBr$_3$ QDs when exposed to air because of fusing and formation of bulk-like domains[34]. Our samples are encapsulated in polymeric matrices to prevent such degradation[35,36]. Moreover, the ability to switch reversibly between superfluorescent and uncoupled spontaneous emission signifies that the quantum confined QDs remain intact. This conclusion is also supported by structural and spectroscopic characterizations (see also SEM micrographs in SI), where no traces of bulk formation or emission are observed. The results are verified by more than 20 independent measurements conducted on different QD superlattices, with the focused electron beam and 4 control measurements with the defocused beam (See Table 1 in SI).

The spectral redshift is explained by a short-range dipole-dipole interaction between excitons. Similar redshifts were measured in molecular J-aggregates and for excitons in CsPbBr$_3$ QDs[2,37]. The central wavelength redshift $\Delta E$ can be estimated by[38]:

$$\Delta E \approx 2|J_c| = \frac{2\mu^2}{4\pi\varepsilon R^3}, \qquad (4)$$

where $\varepsilon$ is the permittivity of the surrounding material (causing screening), $\mu$ is the dipole moment of the exciton, and $R$ is the distance between excitons. For our CsPbBr$_3$ nanocrystals, the dipole moment is $\mu$~7 nm×$e$ [5] and $R$~14 nm, giving a redshift of ~10 nm, similar to the observed redshift.

For comparison, we also analyzed cathodoluminescence from the interaction of a continuous wave (CW) electron beam and the superlattice QDs. In this regime, only the spontaneous emission peak is observed (see Fig. S5), while no superfluorescence is detected. We assign this to the low electron beam current of ~2 nA in CW mode (focused to a beam size of ~50 nm resulting in an electron beam density of ~6×10$^6$ electrons/(s·nm$^2$)) that corresponds to ~0.01 electrons/ps impinging on the superlattice, which is not a sufficient rate to generate superfluorescence. With such a low excitation rate, we do not excite more than one emitter before the dephasing of the excitons (between 50 to 80 ps in the perovskite QDs[6]), and therefore, only spontaneous emission is observed.

**Comparing electron-driven and light-driven superfluorescence**

We expanded our experiment to complement the electron-triggered superfluorescence by light-triggered superfluorescence from the same samples. We excited the sample by a UV laser at cryogenic temperatures (Fig. 3(a)). These experiments emphasized analogies and differences between light-triggered superfluorescence and electron-triggered superfluorescence. The results portrayed in Fig. 3(b) show that emission measured from the samples under optical excitation presents two emission peaks, as reported by Rainò et al.[2] for the same QDs. The blueshifted peak is interpreted as spontaneous emission (uncorrelated QDs) and the redshifted peak as superfluorescent emission (correlated QDs). This spectrum is different from electron-triggered superfluorescence, which allows us to isolate the individual peaks and tune the spectrum from one peak to the other by changing the electron excitation conditions.

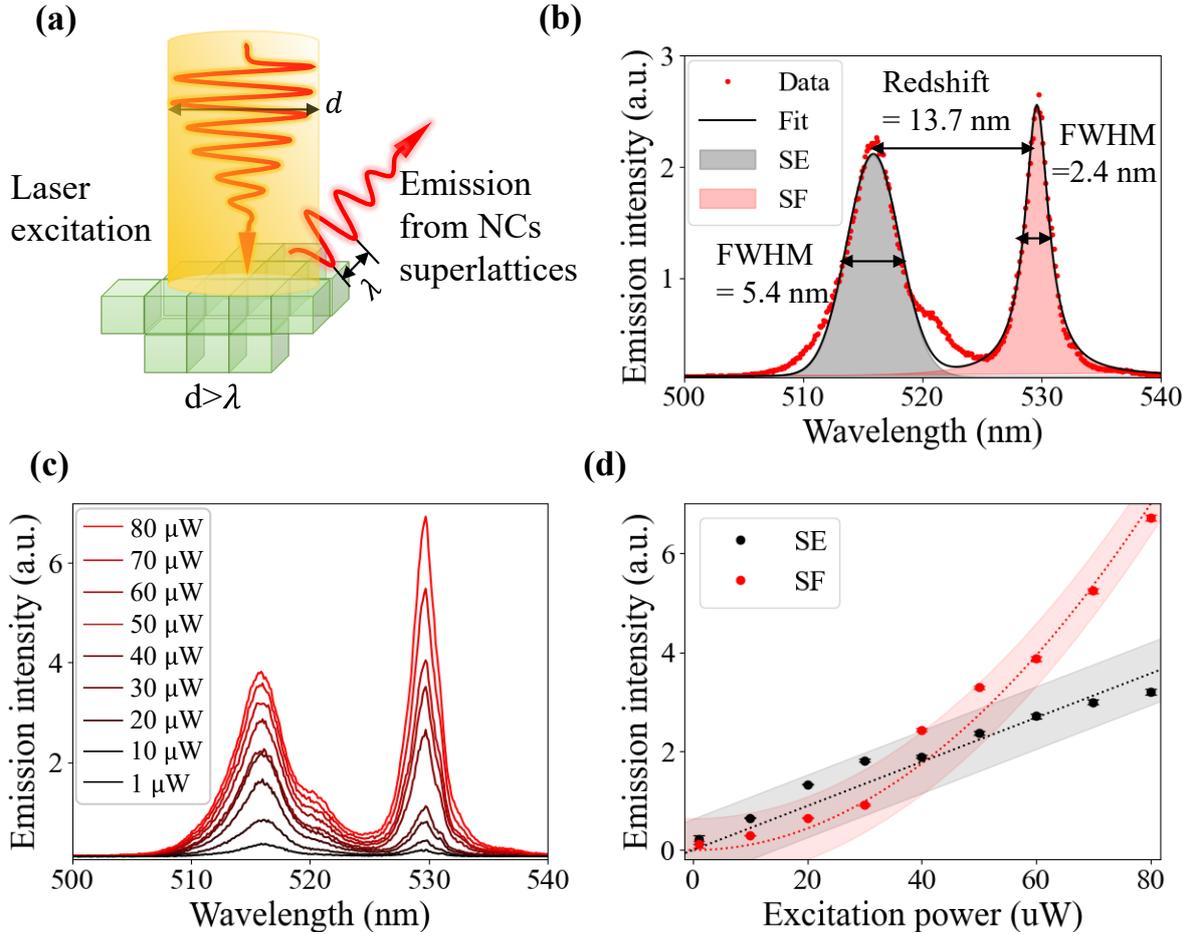

**Fig 3. Validation of superfluorescence from the QDs using optical excitation. (a)** Illustration of conventional superfluorescence triggered by optical excitation. Here, the beam size $d$ is larger than the emitted wavelength $\lambda$ and often overlaps with more than one superlattice. **(b)** Emission spectrum from the optical excitation: we observe spontaneous emission and superfluorescence simultaneously. The superfluorescence peak is located at 530 nm, with a linewidth of 2.4 nm. The spontaneous emission peak is at 515 nm with a linewidth of 5.4 nm. **(c)** Optical spectra for different excitation powers. **(d)** Intensity of the spontaneous emission peaks (black dots) and of the superfluorescence peaks (red dots) as a function of excitation power. The black and red dotted curves fit linear and square functions to the experimental data for spontaneous and superfluorescent emission, respectively.

Optical setups constrained by the diffraction limit lack the spatial resolution to resolve superfluorescent and spontaneous peaks separately, as we demonstrated in the time-resolved cathodoluminescence experiment. Nevertheless, in a power-dependent excitation experiment, we observed the different optical responses of the two peaks. The excitation power varied between 1 and 100 µW (Fig. 3(c)), which corresponds to an excitation difference of 2–200 photons/ps per beam area (with a beam size of ~1 µm, which statistically contains one superlattice of QDs). We can see a functional difference in the intensity of the two emission peaks. The intensity of the blueshifted peak scales linearly with the excitation power, as expected from spontaneous emission. In contrast, the intensity of the redshifted peak scales with a supralinear dependence on the excitation power, as expected from superfluorescent emission (Fig. 3(d)). Both scaling laws agree with the trend predicted by theory[3,9]. For continuous light excitation, superfluorescence was achieved only when the exciting photon flux was sufficient to excite several neighboring QDs faster than the QDs' dephasing time $T_2$=50-80ps [6]. An additional emission peak at 520 nm was observed when the excitation power exceeded 30 µW. This peak is attributed to multi-excitons, bi-excitons, and trions[39,40]. The presence of such multi-excitons explains the deviation from linear scaling for spontaneous emission in Fig. 3(d). Specifically, the central 515 nm peak is reduced at the expense of increased emission at the 520 nm multi-exciton peak.

A comparison of electron- and light-driven experiments thus shows that spontaneous emission and superfluorescence are observed simultaneously in a light-triggered experiment (Fig. 3(b)), while in an electron-triggered experiment we can observe them separately by varying the electron beam size (Fig. 2(g)). Another difference is the linewidth of the spontaneous emission spectral peak. The spectral linewidth of electron-triggered spontaneous emission (defocused electron beam) is three times as broad as its light-driven counterpart (15 nm vs. 5.4 nm). We hypothesize that this broadening stems from the fundamental difference between electron and light interaction with matter: in light–matter interactions, the entire quantized photon energy is transferred in a process of absorption or stimulated emission. In contrast, in electron–matter interactions, energy transfer is not limited to quantized amounts, and a continuous change of energy is possible in the excitation or quenching of the QDs' excitons. Therefore, free-electron excitations below the density threshold for superfluorescence typically create a broader emission spectrum than optical excitations.

The most interesting difference between electron- and light-driven superfluorescence is in the linewidth of the superfluorescence spectral peak. In contrast to the linewidths of spontaneous emission, here we find the electron-driven linewidth to be narrower than the light-driven linewidth (1.9 nm vs. 2.4 nm). We hypothesize that this difference arises from the smaller electron beam size, which can substantially reduce the inhomogeneous broadening, isolating coherent emission with linewidth that is mostly due to intrinsic coherent broadening. This explanation is further supported by the good fit of the Lorentzian line shape (Fig. 2(g)).

**Discussion**

It is valuable to distinguish the free-electron-driven superfluorescence observed in this study from other processes of enhanced free-electron emission. For example, free-electron superradiance[32,41] is a term often used to describe radiation in free-electron lasers and in other processes where the emission is directly from the electrons and enhanced by electron bunching. In another example, electrons impinging specially patterned surfaces (e.g., metamaterials and metasurfaces) excite collective electromagnetic mode[25,27,42,43] that alter the radiation angular distribution. This type of holographic free-electron radiation is emitted from the electron rather than from quantum emitters; i.e., the surface pattern mediates the radiation emission without undergoing electronic transitions as in quantum emitters.

We also distinguish superfluorescence from amplified spontaneous emission, often called superluminescence. The latter arises from *stimulated* emission by an active gain medium (e.g., observed for lead-halide perovskites[44]). In comparison, superfluorescence is purely *spontaneous*, rather than stimulated. The signature characteristics of superfluorescence that we observed – shortened decay time and narrower emission spectrum – do not appear in any of the other processes of enhanced emission. Superfluorescence is uniquely created by the symmetry of the joint wavefunction shared by multiple interacting emitters. Our observed enhancements, much like the broader family of superradiance phenomena, is purely due to transitions between Dicke-superradiant symmetric states.

Our analysis of electron-driven superfluorescence is based on the theory recently proposed in Ref.[45] for the interaction of an electron with a single emitter (free-electron–bound-electron resonance interaction), which has led to substantial theory follow-up works (e.g., Refs.[31,46–49]). We extended the previous works to capture superfluorescence using the symmetric states from the Dicke theory in place of the two-level system model[3]. This theory enables us to emphasize the difference between light- and electron-driven excitations.

**Outlook**

We have demonstrated a new type of superfluorescence: that triggered by pulses of free electrons. i.e., superfluorescent cathodoluminescence. Control of the electron beam size allows us to switch between the distinct emission regimes of spontaneous emission and superfluorescence. We further demonstrated the significant differences between light and electron excitations. A focused free-electron beam drives a pure superfluorescent emission, whereas the light-driven excitation induces a mixture of superfluorescence and spontaneous emission. Our work thus establishes that the underlying nature of superfluorescence is independent of the type of excitation, free-electron or optical.

Superfluorescent cathodoluminescent occurs only when the effective beam area is larger than a single quantum dot and smaller than the correlation distance between quantum emitters. Thus, future experiments could use the electron beam size as a new degree of freedom to extract information about the nature of superfluorescence, revealing the density of correlations with nanometer resolution. This degree of freedom enables us to observe superfluorescent emission separated from any trace of regular spontaneous emission, which has not been possible in conventional optical experiments.

The agreement between experiment and theory motivates us to use our theoretical model as a tool for the design of future fundamental experiments in electron-light-matter interaction. We envision the next steps to utilize electrons for coherent control of multiple correlated emitters at deep subwavelength resolution. This idea, in combination with recent developments in ultrafast electron microscopy [33,47,50–53], suggests that coherent shaping of the electrons' wavefunctions could control coherent effects in superfluorescence that cannot be accessed with optical excitations.

We anticipate that superfluorescence excited by free electrons will have applications in the wider fields of electron microscopy and spectroscopy. For example, electron-triggered superfluorescence can be used to characterize the collective emission from QDs embedded in photonic cavities and waveguides [54] or in integrated photonic on-chip devices [55,56]. Moreover, in bio-related studies, superfluorescent correlated aggregates can be used as extra bright markers for beam-sensitive biological samples in electron microscopy. Such superfluorescent markers would enable deep subwavelength spatial resolution with reduced electron flux for the same acquired signal. A different application could use superfluorescence to improve the sensitivity and response time of free-electron cameras that are based on cathodoluminescence. Such cameras are used in nuclear physics experiments, light intensifiers, electron microscopes, and other electron-based quality inspection systems in the semiconductor industry.

# Supplementary Information

# Free-electron superfluorescence collective optical dynamics at deep-subwavelength resolution

Orr Be'er[1†], Alexey Gorlach[2†], Alina Nagel[1], Reut Shechter[1], Yaniv Kurman[2], Marc Fouchier[3], Ido Kaminer[2], Yehonadav Bekenstein[1*]

*Corresponding author. Email: bekenstein@technion.ac.il

**Supplementary Note 1 - Colloidal CsPbBr$_3$ QDs synthesis**

Materials :

Benzoyl bromide (97%, Aldrich), cesium carbonate (Cs2CO3, 99.9%, Aldrich), lead acetate trihydrate (99.99%, Aldrich), oleic acid (OA, 90%, Aldrich), oleylamine (OLA, 70%, Aldrich), toluene (A.R., Aldrich), Polystyrene (PS) (Aldrich). All chemicals were used as purchased without further purification methods.

Synthesis of CsPbBr$_3$ QDs:

CsPbBr$_3$ NPs are synthesized following a published procedure by Imran et al[1]. With slight modification in temperatures. In a typical synthesis, 16 mg (0.05 mmol) of cesium carbonate, 76 mg (0.2 mmol) of lead acetate trihydrate, 0.3 ml of OA, 1 mL of OLA, and 5 mL of ODE were loaded into a 25 mL 3-neck round-bottom flask, and dried under vacuum for 1 hour at 100 °C. Then, the solution was heated to 170°C under nitrogen, and 0.07 ml (0.6mmol) of benzoyl bromide was swiftly injected. The reaction mixture was immediately cooled down in an ice-water.

Isolation and purification of CsPbBr$_3$ QDs:

5 mL of toluene was added to the crude solutions, and the resulting mixture was centrifuged for 10 minutes at 4000 rpm. The supernatant was discarded, and the precipitate was redispersed in 5 mL of toluene for further use.

Optical and structural characterization of the Colloidal CsPbBr$_3$ QDs synthesis:

The colloidal solution is characterized by absorption-emission spectroscopy and by transmission electron microscopy (TEM - FEI Tecnai G2 T20 S-Twin TEM). The QDs are highly crystalline with an average size of 7.5 nm. The optical properties of the QDs and their crystalline structure are shown in Fig. S3a. The TEM image of the QDs and the distribution of the QDs sizes are shown in Fig. S3b and Fig. S3c.

**Supplementary Note 2 - QDs superlattices in polystyrene thin-film sample preparation**

A Si wafer was cut to 15x15 mm and cleaned by an ultrasonic bath of acetone for 10 minutes, isopropanol for 10 minutes, and HLCP water for 10 minutes then dried by a stream of nitrogen until no visible water droplets appear. The wafer was left for one minute on 110C hot plate and left to cool down for two minutes before the solution spin coating deposition.

QDs in PS solution preparation and deposition:

A 0.36 g of PS was mixed in 9.5 ml of toluene by magnetic stirring. 0.5 ml of the Colloidal CsPbBr$_3$ QDs solution was added to the solution after the PS was fully dissolved. This solution was deposited on the pre-cleaned Si wafer through 1 minute of spin coating at 2000 rpm in a clean room environment.

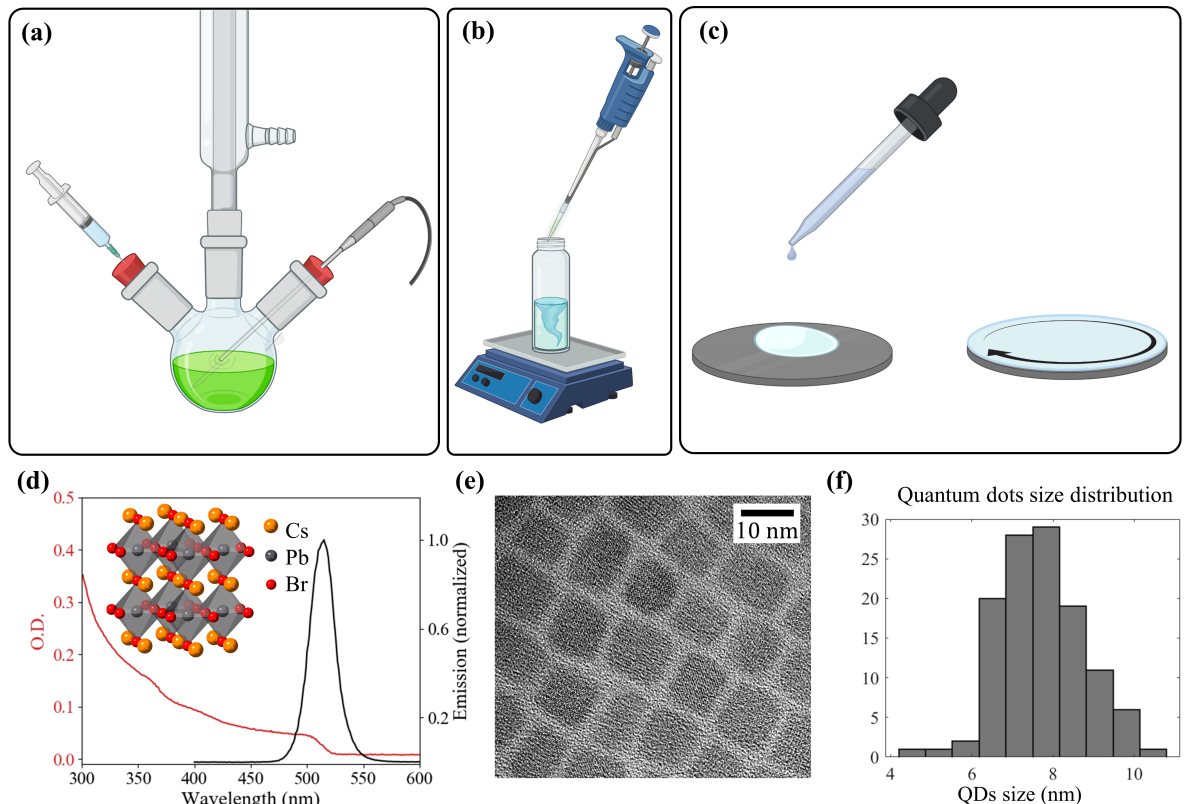

**Fig. S1. (a-c)** Sample preparation schematics **(a)** Synthesis of colloidal CsPbBr3 QDs solution. **(b)** QDs solution mixing in PS solution. **(c)** Deposition of the QDs in PS on a silicon substrate by spin coating **(d)** Optical and structural properties of CsPbBr$_3$ nanocrystals colloidal solution. The red line is the absorption spectrum, and the black line is the emission spectrum excited by 370 nm light. **(e)** TEM tomography of the QDs. The average size of the QDs is ~7.5 nm and consists of many atoms. **(f)** Size distribution of 120 QDs.

**Supplementary Note 3 - Structural characterization**

The sample is characterized by an SEM (Zeiss Ultra-Plus FEG-SEM) and FIB SEM (Dual beam FIB - Helios nano-lab G3 FEI). The results are shown in Figs. 1c, 1d in the main text and Figs S1 and S2. We noticed the superlattices are either at the surface of the polymer or at the polymer and substrate interface.

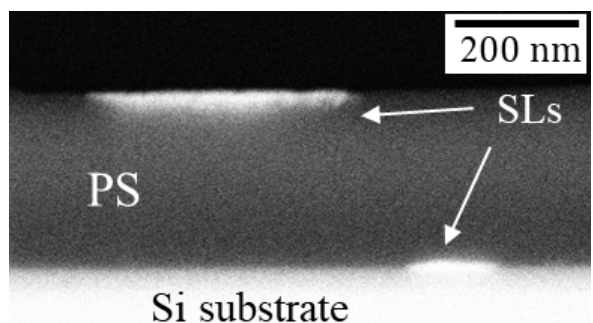

**Fig. S2.** Scanning electron microscope (SEM) micrograph showing a cross-section of the nanocrystal in polystyrene (PS). The sample is cut with a focused ion beam (FEI Helios NanoLab DualBeam G3 UC) and imaged in the same system.

We also noticed that the superlattice's average size is controlled by the QD-to-polymer concentration ratio. We demonstrated that by preparing samples of fixed PS solution and changing the ratio of the colloidal QDs solution from the same batch and toluene while keeping the total volume the same. We observed that the average size of the superlattices size increases with the concentration of the colloidal solution as shown in Fig S2.

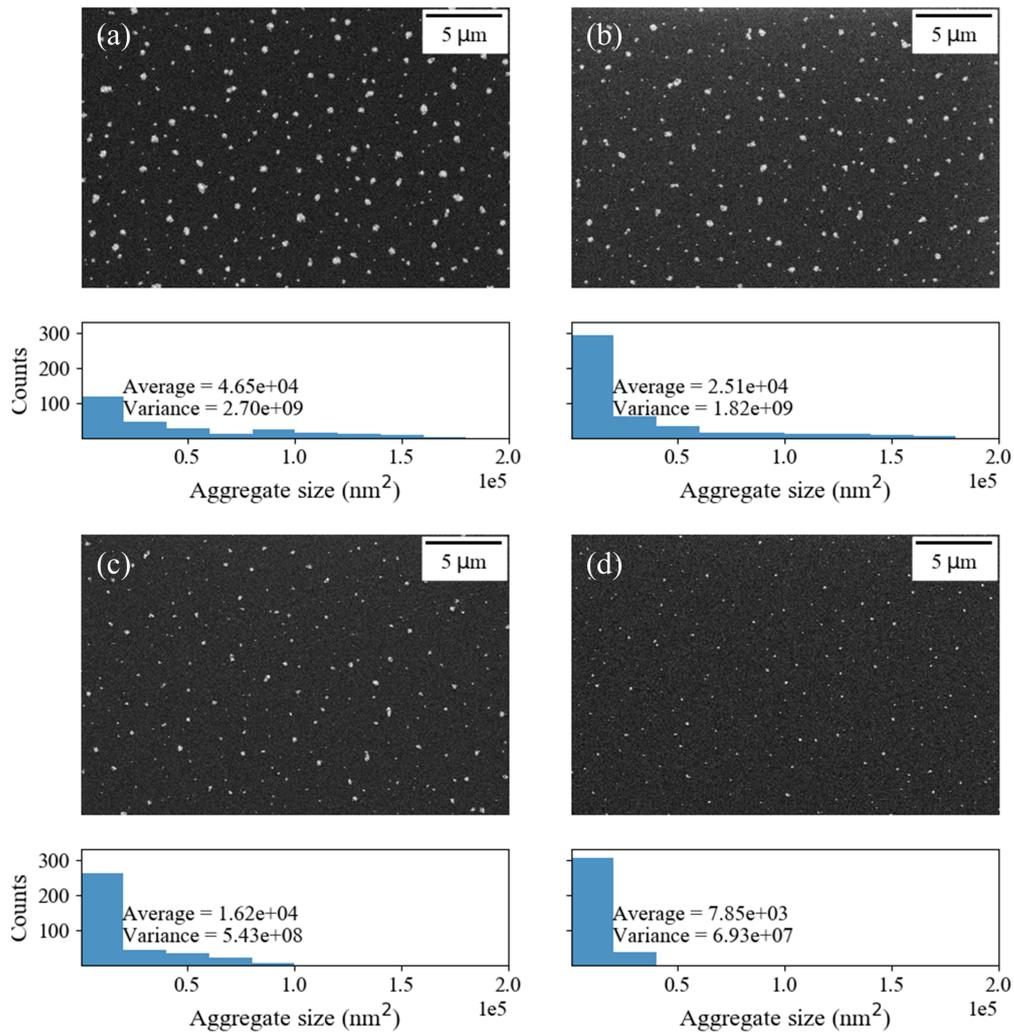

**Fig. S3.** SEM micrographs of samples made using solutions of 4wt% PS and different colloidal solution concentrations **(a)** 1:5 **(b)** 1:10 **(c)** 1:20 **(d)** 1:50.

The excitation of free electron cathodoluminescence (CL)

    All the cathodoluminescence (CL) experiments were conducted on an Allalin Chronos ultrafast scanning electron microscope (USEM) by Attolight. The free-electron excitation experiment takes place inside the USEM vacuum chamber, where the sample is cooled to 10 K to increase exciton coherence times. We use a 3 keV e-beam with a beam current of 2 nA to image and excite the QDs. To generate a pico-second electron pulse, the electron gun was excited by an 80 MHz pulsed laser. The resulting electron pulses contained approximately 12 electrons per pulse. The CL emission is analyzed with a high-speed CDD camera (Andor Newton 920). The CL is analyzed using a streak camera (Hamamatsu C10910) for time-resolved CL (TRCL) measurements, which are shown in Fig. S4.

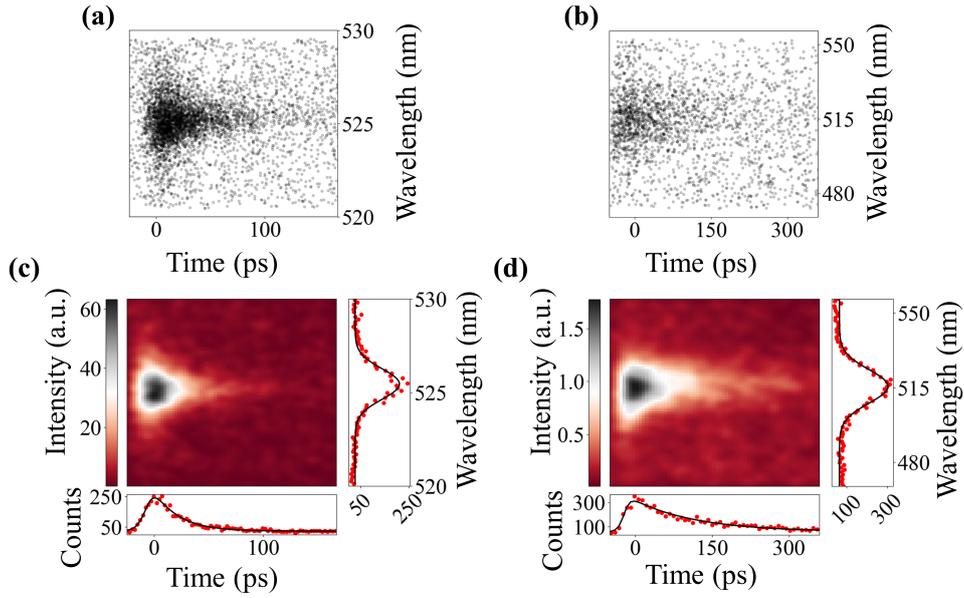

**Fig. S4.** Streak images of time-resolved cathodoluminescence (CL). Note the substantially different wavelength scales and time scales. **(a)** Raw streak camera data for CL was created by a focused electron excitation (10 nm) of a single superlattice. The image is taken across 180 ps along the time axis and 10 nm along the spectral axis from a single superlattice. **(b)** Raw streak camera data for CL was created by a defocused electron excitation (3 μm). The image is taken across 410 ps along the time axis and 90 nm along the spectral axis. **(c)** and **(d)** The raw data was processed using Gaussian kernel density estimation (KDE)[2]. The right insets present the corresponding emission spectra by integrating over the time axis. These spectra are fitted to a Lorentzian in (c) and a Gaussian in (d). The bottom insets present the time-dependent emission by integrating over the wavelength axis. The values of the decay times were extracted by fitting them to an exponent.

Our focused electron pulse measurements were repeated 22 times, each measuring a different superlattice. The defocused beam measurements were repeated 4 times, covering different areas. The result analysis is summarized in table S1

|  |  |  | Gaussian fitting |  | Lorentzian fitting |  |
|---|---|---|---|---|---|---|
| Samples | Emission wavelength | Emission decay time (ps) | Average FWHM (nm) | Average $\chi^2$ (nm) | Average FWHM (nm) | Average $\chi^2$ (nm) |
| 4 – uncoupled | $515.6 \pm 0.4$ | $102 \pm 2$ | $17 \pm 2$ | $14.4 \cdot 10^3$ | $18 \pm 3$ | $21.1 \cdot 10^4$ |
| 22 – coupled | $525.4 \pm 0.1$ | $21 \pm 3$ | $1.6 \pm 1$ | $71.4 \cdot 10^3$ | $1.9 \pm 0.2$ | $57.3 \cdot 10^3$ |

**Table S1**. Summary of 26 time-resolved CL experiments and the appropriate temporal and spectral statistical analysis. Each row summarizes 4 (22) measurements carried out using a defocused (focused) electron beam excitation, triggering emission from uncoupled (coupled) quantum dots (QDs) in superlattices. These measurements show the 5-fold reduction of the decay time and a 10 nm redshift of the emission wavelength for coupled QDs compared with uncoupled QDs. The spectral analysis shows that the spectral shape of emission from uncoupled QDs fits better to a Gaussian, while the spectral shape of emission from coupled QDs fits better to a Lorentzian.

Continuous-wave (CW) CL from QDs superlattices

We compare the measurements of TRCL obtained using electron pulses with conventional CL obtained using regular, i.e., CW electron beams. Fig. S5 a hyperspectral image of the sample was taken in CW mode. Panels (a) and (b) show a direct correlation between the superlattices' SEM secondary electron image and the CL image. Each spectral point contains a single broad peak (FWHM>15 nm) where no spectral traces of superfluorescence (i.e., peaks with FWHM<5) appeared. The e-beam current here is 2 nA and the beam spot size is ~50 nm.

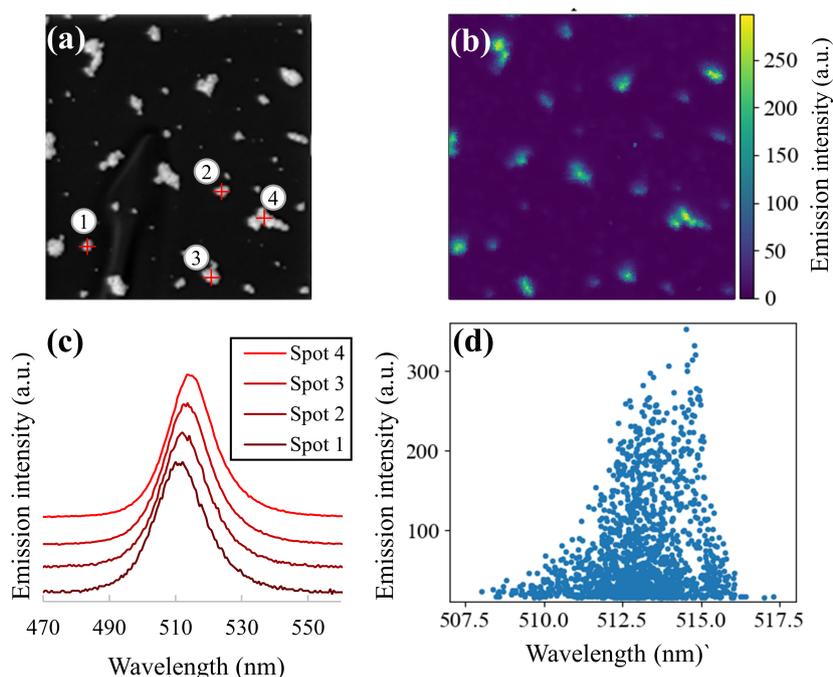

**Fig. S5.** Hyperspectral analysis of the QD superlattices. **(a)** SEM image (of secondary electrons) of the sample, where the white spots are the QD superlattices **(b)** A map of emission intensity from the sample. The emission from the superlattices is prominent over the background. **(c)** The normalized emission spectrum from points 1-4 is marked on (a). Along the entire sample, no spectral traces of superfluorescence are observed (i.e., redshifted, narrow peak). **(d)** Correlation between the emission intensity and the peak wavelength. The variation of peak wavelength may occur due to the variation of the size and quantum confinement of the QDs and the redshift in high intensities due to the inner-filtering effect.

Stability of $CsPbBr_3$ QDs under electron beam irradiation

Material structural stability is of great concern in electron beam characterization. Beam damage is known to extensively affect both halide perovskites[3,4] and lead-free perovskites[5]. During the pulse excitation experiment (discussed in the main text) we observed a decrease of the signal with no change in the emission spectrum. We attribute this degradation to carbon contamination that could arise from the polymer matrix. However, we expect that in the CW regime there is a significant degradation. We perform a sequence of measurements over a period of 6 minutes to estimate the degradation of the sample under CW e-beam excitation. The electron beam is focused to ~50 nm with a current of ~2 nA. In this regime, only the spontaneous emission peak is observed Fig. S6 presents CL spectra and their gradual evolution during the experiment. The emission peak decays to half the intensity after exposure of ~20 s and continues to slowly degrade. The decay is not uniform and reveals an underlying weaker emission presented as a shoulder at 527 nm. We attribute this to the fusing of QDs and the formation of bulk $CsPbBr_3$. Such emission agrees with previous work on bulk $CsPbBr_3$ with the same CL microscope[6].

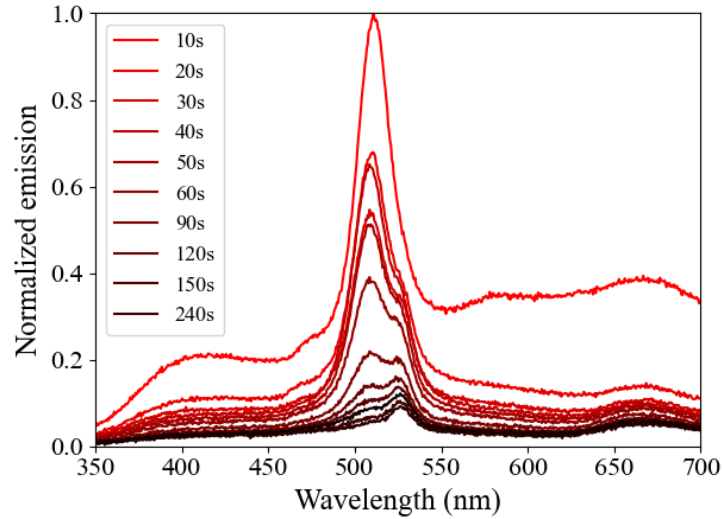

**Fig. S6.** The evolution of the CL emission from the sample over time, under a continuous-wave (CW) e-beam excitation, showing the timescale under which degradation of the sample occurs.

Optical measurements

To compare with the CL measurements, we also performed optical measurements on the same samples, using a 405 nm CW diode laser. The results are depicted in Fig. 4a of the main text. The laser beam is focused using a 20x objective with NA of 0.29 and the working distance of 31 mm, reaching a laser spot diameter of 880 nm on the cooled sample (~10 K). The spectrum of the emission is detected with a TRIAX 552 spectrometer equipped with a cooled CCD (Roper Scientific).

**Supplemenary Note 4 - Theoretical description of free electrons superfluorescence**

The following supplementary section provides a detailed discussion of the theoretical model used to explain the observed superfluorescence triggered by free electrons. We found that using a two-level system (TLS) model to describe each QD's exciton is a good approximation that captures the main features observed in our experiment. We construct a theoretical model that describes the interaction of electrons with multiple TLS emitters, where each electron can either excite or quench the energy of the excitons' collective. We show that collective emission (i.e., superfluorescence) is indeed possible in this case.

A typical electron interaction can transfer more energy than required for a direct TLS excitation. For example, the electron usually excites a bulk plasmon, which then excites one or more electron-hole pairs that relax to create the effective TLS excitation[7]. Therefore, our theory cannot account for the electron energy loss. The theory accounts for the excitation of the multi-TLS system of QDs' excitons, which creates the necessary initial condition for the superfluorescent emission dynamics.

Interaction of a single electron with a single two-level system

In this section, we show how the quantum theory of interaction between a free electron and a two-level system corresponds to the known theory of free-electron-atom interaction (Fig. S7), described back in 1930s by Bethe[8] and can be found in the textbooks[9]. This theory was recently revisited in more modern contexts, which could be applicable for experiments in electron microscopes, under the name free-electron bound-electron resonant interaction (FEBERI) [10]. Quantum mechanical treatments [11,12] followed the original semiclassical theory.

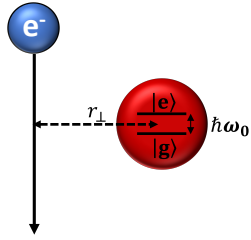

**Fig. S7.** Interaction of an energetic electron with a single two-level system (TLS).

Using the language of quantum optics, the scattering matrix of the interaction between a free electron and a single TLS can be calculated according to the following formula [10–12]:

$$S = e^{-i(gb\sigma^+ + g^*b^\dagger \sigma^-)}, \quad (S1)$$

where $b, b^\dagger$ are the electron energy shift operators ($b$ reduces the electron energy by $\hbar\omega_0$, while $b^\dagger$ increases it by $\hbar\omega_0$), and $\sigma^-, \sigma^+$ are the ladder operators of a TLS. The interaction constant $g$ is:

$$g(r_\perp) = \frac{e\omega_0}{2\pi\varepsilon_0 \hbar v^2}\left(d_\perp K_1\left(\frac{2\pi r_\perp}{\lambda_0 v/c}\right) + d_\parallel K_0\left(\frac{2\pi r_\perp}{\lambda_0 v/c}\right)\right). \quad (S2)$$

$e$ is an elementary charge, $v$ is the speed of the electron, $d_\perp$ and $d_\parallel$ are the projections of the transition dipole moment of the quantum dot on the direction perpendicular and parallel to the electron motion to the motion, respectively, $r_\perp$ is the distance between the electron's trajectory and the position of the TLS (Fig. 1S). Eq. (S2) can be significantly simplified in case the effective distance to the exciton $r_\perp$ is small:

$$\frac{2\pi r_\perp}{\lambda_0 v/c} \ll 1. \quad (S3)$$

For example, for electron kinetic energy of 3 keV and $\lambda_0 = 500$ nm, we get the condition of Eq. (S3) to be equivalent to a distance of $r_\perp \ll 80$ nm. Under this condition, Eq. (S2) can be simplified to:

$$g(r_\perp) \approx \frac{ed_\perp}{2\pi\varepsilon_0 \hbar v\, r_\perp}. \quad (S4)$$

Then the probability to excite a QD by a single electron is

$$P_e = |\langle e| \otimes \langle E - \hbar\omega_0|S|E\rangle \otimes |g\rangle|^2,$$

where $|E\rangle$ and $|E - \hbar\omega_0\rangle$ are the electron states with energies $E$ and $E - \hbar\omega_0$. $|g\rangle$ and $|e\rangle$ are the ground and excited states of the TLS. Substituting Eq. (S1), we get the following expression for the probability of the excitation:

$$P_e = \sin^2|g(r_\perp)|. \quad (S5)$$

The probability to excite the TLS in the approximation of $|g| \ll 1$ equals to:

$$P_e = \sin^2|g(r_\perp)| \approx |g(r_\perp)|^2 = \frac{4e^2|d_\perp|^2}{(4\pi\varepsilon_0)^2 \hbar^2 v^2\, r_\perp^2}. \quad (S6)$$

Eq. (S6) is a well-known equation for the excitation of an atom by a charged particle[9].

We note that for electron interactions with solid-state systems, there are additional mechanisms that compete with this direct interaction, such as the excitation of a bulk plasmon. The unique aspect of the interaction described here is that it maintains coherence throughout the interaction. In contrast, an excitation of the TLS through other mechanisms, such as through bulk plasmon excitation, will not maintain coherence, since part of the electron energy will go to other channels. As we show in the sections below, maintaining coherence is necessary to create the effect of superradiance, but not to create the effect of superfluorescence. In our experiment, we do not know whether the electron excitation is direct (as in this section) or indirect, and thus we can only claim to observe superfluorescence, rather than superradiance. It remains an open challenge for future experiments to provide the direct experimental evidence necessary for a claim of electron-driven superradiance. This for example could be done using a combined CL and electron energy loss spectroscopy (EELS)[13], where the amount of energy lost by the electron could be correlated with the energy transferred to CL.

Interaction of multiple electrons with a single two-level system

In this section, we show that free electrons cannot fully excite a TLS. Specifically, free electrons can only create TLS states with excited level probabilities of up to one half. According to Eq. (S5), after the interaction with $N_e$ consequent electrons, the probability of the TLS to be excited equals to

$$P_{N_e} = \sin^2 g \, \frac{(1-\cos^{N_e} 2g)}{1-\cos 2g} \leq 1/2. \tag{S7}$$

Fig. S8 presents $P_{N_e}$ as the function of $N_e$ for a fixed $g$. The limitation on the TLS excitation can be explained simply: when the TLS is excited to less than on half (i.e., $P_{N_e} \leq 1/2$) then the electron on average transfers its energy to the TLS; when the TLS is excited to more than on half (i.e., $P_{N_e} \geq 1/2$) then the electron on average gains energy from the TLS. So, if the initial TLS was in the ground state ($P_{N_e} = 0$), after its interaction with any number of electrons, it will remain at $P_{N_e} \leq 1/2$. As we show below, this limit prevents us from reaching part of the effects of superradiance but does not prohibit creating the effects of superfluorescence. We note that this limit can be overcome by pre-shaping the electron wavefunction, as was predicted theoretically[14].

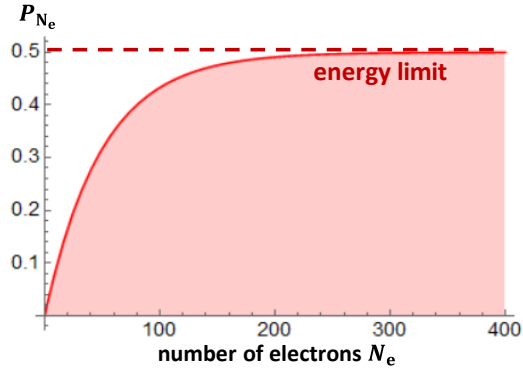

**Fig. S8.** Excitation probability of a TLS as a function of the number of electrons in the beam. The probability of excitation is bounded by 50%. The simulation here assumes interaction constant $g = 0.1$.

Incoherent interaction of electrons with multiple two-level systems

In this section, we discuss the incoherent interaction of free electrons with multiple TLS. This theory usually describes electron interactions in thick materials (containing at least a few dozens of atomic layers), and is observed in other experimental cases of cathodoluminescence (CL) with not correlated emission. We show that this type of interaction is captured by our theory. Let us consider a material where TLS are distributed uniformly inside the sample (Fig. S9).

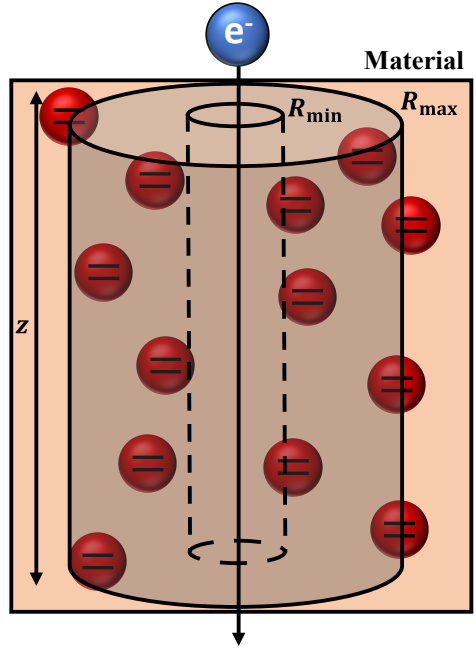

**Fig. S9.** Interaction of an energetic electron with a sample containing a uniform distribution of TLSs.

According to section S1, under the approximation of Eq. (S3), we can write the probability of a single TLS excitation as $P_e = \frac{4e^2 |d_\perp|^2}{(4\pi\varepsilon_0)^2 \hbar^2 v^2 \, r_\perp^2}$. When considering only this interaction mechanism, the average energy loss of an electron after the interaction with a single TLS is thus:

$$\langle E \rangle = P_e \cdot \hbar\omega_0 = \frac{4e^2 |d_\perp|^2 \omega_0}{(4\pi\varepsilon_0)^2 \hbar v^2 \, r_\perp^2}.$$

Now consider a cylindrical volume (Fig. S9) with an internal radius $R_{min}$, external radius $R_{max}$, and length $z$. The total electron energy loss due to the spatial distribution of the TLS in this cylinder is

$$E = n \cdot z \cdot \frac{8\pi e^2 |d_\perp|^2 \omega_0}{(4\pi\varepsilon_0)^2 \hbar v^2} \int_{R_{min}}^{R_{max}} \frac{dr}{r},$$

where $n$ is the number of TLS per unit volume. We can translate this formula to the loss of electron per unit length:

$$\frac{dE}{dz} = \frac{8\pi e^2 n |d_\perp|^2 \omega_0}{(4\pi\varepsilon_0)^2 \hbar v^2} \ln \frac{R_{max}}{R_{min}}. \tag{S8}$$

The minimum radius is connected with the effective size of an exciton (i.e., excited TLS) $R_{min} = a$, while the large radius is connected with the approximation of Eq. (S3): $R_{max} \approx \frac{v}{\omega_0}$. Thus, we get:

$$\frac{dE}{dz} = \frac{4\pi n e^2 |d_\perp|^2 \omega_0}{(4\pi\varepsilon_0)^2 \hbar v^2} \ln \left( \frac{v^2}{\omega_0^2 a^2} \right). \tag{S9}$$

Eq. (S9) represents the non-relativistic Bethe-Bloch formula in the case of the free electron [9]. With the proper estimation of the dipole moment, exciton's radius, and transition frequency, we arrive at the conventional form of the Bethe-Bloch equation:

$$\frac{dE}{dz} = \frac{4\pi n}{m_e v^2} \left( \frac{e^2}{4\pi\varepsilon_0} \right)^2 \ln \left[ \frac{2m_e v^2}{I} \right], \tag{S10}$$

where $I$ is the ionization potential and $m_e$ is the electron mass (appearing by substituting formulas for the exciton's radius and the effective dipole). This equation is also the conventional result found in the literature on incoherent CL[15].

Coherent interaction of free electrons with matter

This section considers a novel regime of interaction between free electrons and quantum emitters. This interaction regime is completely different from the conventional cases described in the previous sections, and we find it to be the one relevant to our experiment. Reaching this novel interaction regime requires satisfying several strict conditions. For example, we work at very low temperatures to main coherence between the emitters for the entire duration of their emission. We develop a simple qualitative model that can explain the main features of the observed superfluorescence.

We assume a dense cluster of emitters (e.g., QDs) that satisfies two conditions on the parameters defined in Fig. S10:

$$\begin{cases} d \ll r_\perp \\ d \ll \lambda_0 \end{cases}. \quad (S11)$$

We should note that such a situation is completely different from what we considered in the previous section (Fig. S9). While in the previous section we considered a *uniform* distribution of emitters in the sample, here we consider an opposite situation – a dense cluster of *concentrated* emitters, where we can treat all the emitters as being approximately at the same point. Interestingly, this case can be described by a direct generalization of the formalism of Eq. (S1).

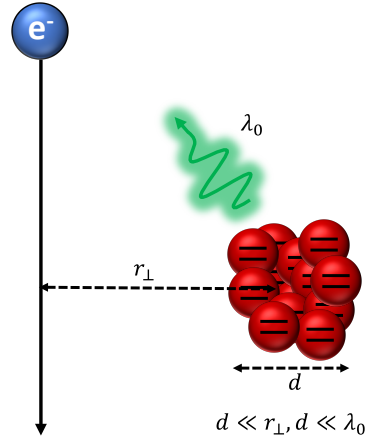

**Fig. S10.** Interaction of an energetic electron with a dense cluster of emitters

The scattering matrix for the emitters can now be described by the replacing the individual emitter operators $\sigma^\pm$ in Eq. (S1) that describe a spin half system, by the operators $S^\pm$ of a general (higher than one half) spin state:

$$S = e^{-i(gbS^+ + g^* b^\dagger S^-)}. \quad (S12)$$

These operators $S^\pm$ describe the raising and lowering operations on the ladder of symmetric states, which can be written as the sum $S^\pm = \sum_i \sigma_i^\pm$ over all the emitters. The Hilbert state of $N$ indistinguishable emitters is described in terms of the symmetric states defined as

$$\begin{cases} |N\rangle = |eee \dots ee\rangle, \\ |N-1\rangle = \sqrt{\frac{1}{N}}(|gee \dots ee\rangle + |ege \dots ee\rangle + \dots + |eee \dots eg\rangle), \\ \dots \\ |0\rangle = |ggg \dots gg\rangle. \end{cases} \quad (S13)$$

where $|m\rangle$ is the symmetric state with $m$ excited emitters. The matrix element of the scattering matrix in Eq. (S12) can be calculated in the following way:

$$\langle m|S|n\rangle = b^{n-m} S_{mn}, \qquad (S14)$$

with $S_{mn}$ being

$$S_{mn} = \sqrt{m!\,n!\,(N-n)!\,(N-m)!} \sum_{k=0}^{n} \frac{(-1)^k (\cos|g|)^{N-m+n-2k} (i\sin|g|)^{m-n} (\sin|g|)^{2k}}{k!\,(n-k)!\,(m-n+k)!\,(N-m-k)!}.$$

The initial state of the joint system of electron and cluster of emitters is

$$\rho^{(i)} = |0\rangle\langle 0| \otimes |E\rangle\langle E|, \qquad (S15)$$

$|0\rangle$ is the state of the cluster with all of the TLS are in the ground state, and $|E\rangle$ is the state of the electron with energy $E$. The density matrix of the TLS after the interaction with electrons is

$$\rho_{\text{em}}^{(f)} = \text{Tr}_e\left(S\rho^{(i)}S^\dagger\right).$$

$\text{Tr}_e$ is the trace over the electron degrees of freedom, which enables to calculate the final state of the emitters

$$\rho_{\text{em}}^{(f)} = \sum_n P_n |n\rangle\langle n| \qquad (S16)$$

that contains the probabilities $P_n$ to $n$ excitons:

$$P_n = \frac{N!}{(N-n)!\,n!} P_e^n (1-P_e)^{N-n}, \qquad (S17)$$

with $P_e$ being the probability to excite a single TLS defined by Eq. (S3). Eq. (S17) represents the binomial distribution. In the case of $P_e \ll 1$ and $N \gg 1$ we get:

$$P_n \approx \frac{e^{-\lambda}}{n!} \lambda^n, \qquad (S18)$$

where $\lambda = P_e N = |g|^2 N$. We note that in this limit, Eq. (S18) represents the conventional probability of multi-excitation by a single electron, corresponding to results often observed in experiments of electron energy loss spectroscopy[16]. As we show below, this theory provides a good quantitative model for our experiment. Even for larger cluster sizes and larger interaction areas that go slightly beyond the conditions of Eq. (S11), we find the theory to still provide a good qualitative explanation for the features observed in our experiment.

Superradiance effects triggered by free electrons

The difference between our work and most previous works on CL is the combination of: (1) the large dipole moment of the emitters ($\approx 7$ nm $\cdot e$, where $e$ is an elementary charge); (2) close distance between emitters (approaching the conditions of Eq. (S11)); and (3) long coherence times due to the low temperature that enables to maintain correlations between the emitters. Satisfying all these conditions is what enables observing for the first time superfluorescence triggered by free electrons. In comparison, previous papers such as[7] observed CL from closely packed semiconductor nanocrystals but did not observe superradiance phenomena, since the temperature was not low enough to support sufficient coherence between emitters.

To describe the dynamics of emission in the experiment, we build a simple model which combines the results of the previous section with the Dicke superradiance theory[17]. According to the results of the previous section, after interaction with a single electron, the density matrix of the emitters is described by Eq. (S16). Let us now consider the interaction with multiple consequent electrons with a time delay $\tau$ between each pair as depicted in Fig. S11 (this delay $\tau$ will be later taken as the mean distance between electrons in the distribution of random electron arrival times).

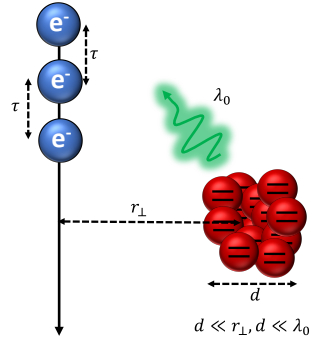

**Fig. S11.** Interaction of multiple energetic electrons with a dense cluster of emitters

The state of the cluster of emitters after the interaction with a single electron is described by

$$\rho_{\text{em}\,mm}^{(f)} = \langle m|\text{Tr}_e(S\rho^{(i)}S^\dagger)|m\rangle = \sum_n |S_{mn}|^2 \rho_{\text{em}\,nn}^{(i)}. \tag{S19}$$

Eq. (S19) can be written in matrix form if we define the column of diagonal elements of emitters density matrix:

$$\vec{\rho}_{\text{em}} = \begin{pmatrix} \rho_0 \\ \rho_1 \\ \ldots \\ \rho_N \end{pmatrix}, \text{ where } \rho_n = \langle n|\hat{\rho}_{\text{em}}|n\rangle.$$

Then Eq. (S19) has the following form:

$$\vec{\rho}_{\text{em}}^{(f)} = \hat{P}\vec{\rho}_{\text{em}}^{(i)}, \tag{S20}$$

where $\hat{P}_{mn} = |S_{mn}|^2$. After the excitation by the first electron and before the second one, the collective emission takes place if all the necessary conditions are satisfied[17]:

$$\frac{d\rho_{\text{em}\,mm}}{dt} = -\Gamma m(N - m + 1)\rho_{\text{em}\,mm} + \Gamma(m+1)(N-m)\rho_{\text{em}(m+1)(m+1)}. \tag{S21}$$

Here $\Gamma$ is the decay rate of a single emitter. Eq. (S21) can be written in matrix form:

$$\frac{d\vec{\rho}_{\text{em}}}{dt} = \widehat{M}\vec{\rho}_{\text{em}}, \tag{S22}$$

with $\widehat{M}_{mn} = -\Gamma m(N-m+1)\delta_{mn} + \Gamma(m+1)(N-m)\delta_{(m+1)n}$. The solution of Eq. (S22) is

$$\vec{\rho}_{\text{em}}(t) = e^{\widehat{M}t}\vec{\rho}_{\text{em}}(0), \tag{S23}$$

where $e^{\widehat{M}t}$ is the matrix exponent. If we have the pulse of $N_e$ electrons, then the density matrix as the function of time has the following form:

$$\vec{\rho}_{\text{em}}(t) = e^{\widehat{M}t} \cdot P(e^{\widehat{M}\tau}P)^{N_e-1}\vec{\rho}_{\text{em}}(0). \tag{S24}$$

The intensity of emission in this process is described by:

$$I(t) = -\hbar\omega_0 \sum_m \dot{\rho}_{\text{em}\,mm}(t) = -\hbar\omega_0 \sum_m (\widehat{M}\vec{\rho}_{\text{em}})_{mm}. \tag{S25}$$

We can write the density matrix as a column $\vec{\rho}_{em}$ because off-diagonal terms of the emitter's density matrix are zero during the process. The plot of Eq. (S25) is shown in Fig. S12. Eq. (S25) was also used to model the intensity in Fig. 1c and Fig. 3f in the main text.

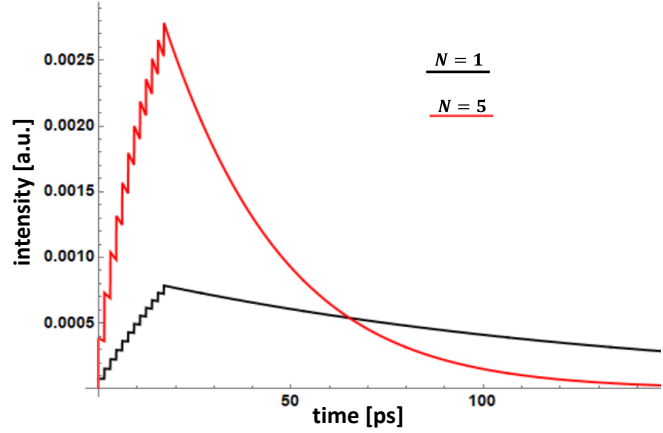

**Fig. S12.** The intensity of emission from a cluster of $N$ emitters triggered by free electrons. We choose the simulation parameters close to the experimental condition: interaction coupling of $g = 0.1$, number of electrons of $N_e = 12$, single-emitter emission rate of $\Gamma = 7.8 \cdot 10^{-3}$ ps$^{-1}$, and delay $\tau$ between consequent electrons of 1.5 ps.

According to Eq. (S19), the electron excitation is instantaneous in time, which is a good approximation since the electron pulse duration (~1 ps) is much smaller than the superradiance emission time (~20 ps in the experiment). The "steps" in Fig. S12 arise from the discrete nature of electrons and their instantaneous interactions. These features should be averaged out by the distribution of electron arrival times. In practice, the dominant averaging mechanism is the finite response time of the measurement device, which averages over these temporal features, resulting in Fig. S13 and in Fig. 1b and Fig. 3b in the main text. All these effects can be modeled by the convolution of the intensity as a function of time $I(t)$ with a Gaussian of a finite width $\tau_0$:

$$\langle I(t) \rangle = \frac{1}{\tau_0 \sqrt{\pi}} \int_{-\infty}^{\infty} I(\tau) e^{-\frac{(\tau-t)^2}{\tau_0^2}} d\tau, \qquad (S26)$$

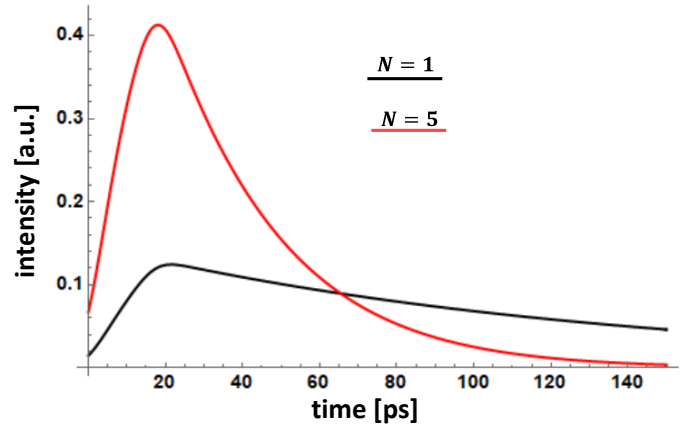

**Fig. S13.** Time-dependent intensity of emission from a cluster of $N$ emitters triggered by free electrons for the same parameters as in Fig. S4, with an averaging Gaussian width of $\tau_0 = 5$ ps.

Fig. S13 shows how our theory can match the temporal dynamics in the experiments. However, there is one assumption in constructing our theory that is not precisely satisfied by most of the clusters in our experiment: The first condition in Eq. (S11) – that the size of the cluster is much smaller than the distance between the electron and the cluster ($d \ll r_\perp$) – is not satisfied. Therefore, the theory we developed may be sufficient for the temporal dependence (as in Fig. S13), but cannot precisely capture the spatial effects, which relate to the size of the electron beam and to the size of the cluster. To build a more accurate theory that can capture the spatial effects, we relax the condition $d \ll r_\perp$ while still maintaining $d \ll \lambda_0$ (where $\lambda_0$ is the wavelength of the emitted light). These relaxed conditions can still support a type of superradiance, as we discuss in the next section.

The spatial dependence of superfluorescence triggered by free electrons

In this section, we consider the interaction of a finite-width electron beam with emitters that are spatially distributed in a square lattice, similar to the experimental configuration. In this case, the effective area of electron excitation is smaller than the size of the cluster. Therefore, the strength of the interaction is different for different emitters in the cluster. This condition is what makes free-electron-driven superfluorescence so different than the more conventional light-driven superfluorescence. In the light-driven case, all the emitters within a certain area experience the same excitation strength (this area is controlled by beam spot size, which is always larger or equal to the wavelength). In this case of light-driven superfluorescence, the theory developed in the previous sections is in good agreement with experiments. However, electron-driven superfluorescence requires a more advanced theory, because the small size of the electron probe causes the excitation strength to vary substantially between different emitters. This spatial dependence provides another degree of freedom for research of superfluorescence, which could enable probing collective dynamics with resolution comparable and potentially smaller than the correlation distance.

The formalism in this section is developed to capture the spatial effects of electron-driven superfluorescence, and its dependence on the electron beam spot size. We consider a single electron that interacts with a square lattice of emitters, with a distance between them of 12 nm, as depicted in Fig. S14a. According to Eq. (S3), the probability of excitation is:

$$P_\text{e} = \sin^2|g(r)|.$$

Fig. S14a schematically shows the square lattice of emitters and the electron, which interacts with the lattice. Fig. S14b shows that a single electron cannot excite more than a single emitter since the effective interaction distance is smaller than a single QD emitter for the values of the parameters in our experiment ($g = 0.1$, distance between emitters $=12$ nm, $v/c = 0.1$).

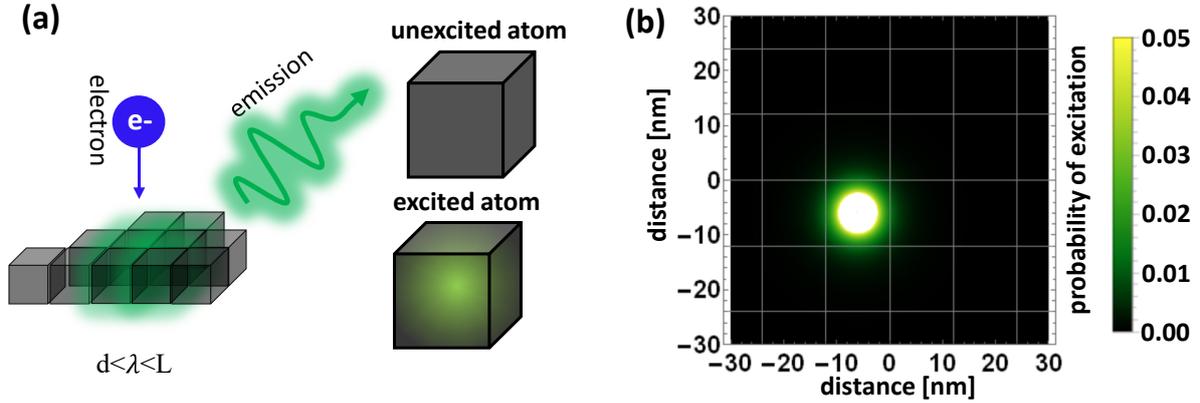

**Fig. S14. (a)** An electron excites a square lattice of emitters (a single layer), with lattice period of 12 nm. **(b)** The excitation probability of an emitter $P_\text{e}$ as a function of distance for the interaction constant $g = 0.1$, showing that a single electron cannot excite more than one emitter for such geometry (thus it cannot create superfluorescence).

We consider a Gaussian electron beam of $N_\text{e}$ electrons, as depicted in Fig. S15a, which is described by the transverse probability density

$$\rho(r) = \frac{N_\text{e}}{2\pi\sigma^2} e^{-\frac{r^2}{2\sigma^2}}, \tag{S27}$$

where $\sigma$ is the spot size of the beam. Now, let us calculate the excitation by a Gaussian beam of many electrons. In this case, the electrons-material interaction constant changes to the convolution of Eq. (S27) and Eq. (S3):

$$P_{\text{beam}}(r) = \int_0^\infty \rho(|\mathbf{r} - \mathbf{r_0}|) \cdot \sin^2|g(\mathbf{r_0})| \, d^2\mathbf{r_0}. \tag{S28}$$

$\rho(|\mathbf{r} - \mathbf{r_0}|)$ is the transverse probability density of the electron beam described by Eq. (S27). $g(\mathbf{r_0})$ is the interaction constant for a single electron described by Eq. (S3). Eq. (S28) can be simplified to

$$P_{\text{beam}}(r) = \frac{N_e}{\sigma^2} e^{-\frac{r^2}{2\sigma^2}} \int_0^\infty e^{-\frac{r_0^2}{2\sigma^2}} I_0\left(\frac{rr_0}{\sigma^2}\right) \cdot \sin^2|g(r_0)| \, r_0 dr_0, \tag{S29}$$

with $I_0\left(\frac{rr_0}{\sigma^2}\right)$ being the modified Bessel function of the first kind.

Fig. S15a depicts the electron beam's interaction with the sample. Fig S15b shows the excitation by the electron beam (i.e., $P_{\text{beam}}(r)$). If just one electron is used for the excitation, it cannot excite superradiance because it excites an area smaller than the size of an emitter. i.e., despite the electron kinetic energy being sufficient to excite multiple emitters, this energy can only be transferred to a small area, and thus only a single emitter can be excited, resulting in no superfluorescence. In contrast, an electron beam with a spot size $\sigma$ larger than the size of the emitter ($\sigma \gg 10$ nm), consisting of many electrons ($N_e \gg 1$), can excite multiple emitters (Fig. S15b) and in this way initiate superfluorescent emission. In the extreme case of increasing the beam size further ($\sigma > 100$ nm), while the number of electrons inside the beam stays constant, we find no superfluorescence due to the low density of excitations, as shown below.

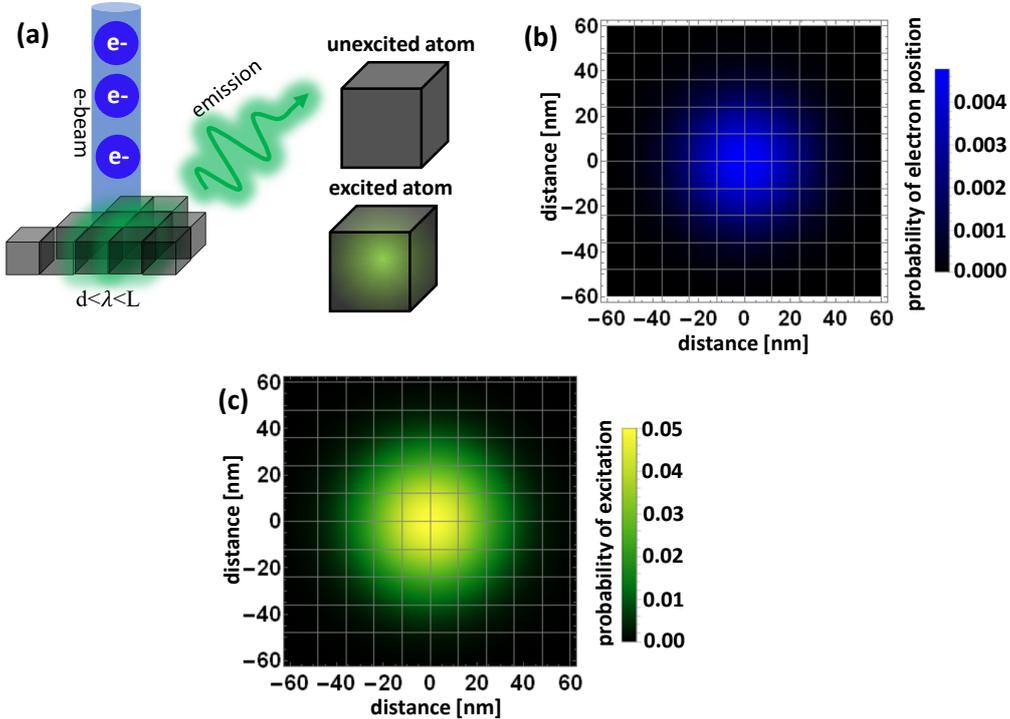

**Fig. S15. (a)** Electron beam of 12 electrons with radius $\sigma = 20$ nm excites a square lattice of emitters (a single layer), with a period of 12 nm. **(b)** The electron beam density. **(c)** The excitation probability $P_{\text{beam}}$ of emitters as a function of emitter location, showing that a beam of multiple electrons can excite more than one emitter (and thus create superfluorescence).

The total number of exited emitters can be calculated by integration over the square lattice divided by the area occupied by a single emitter:

$$\langle N_{\text{exc}} \rangle = \frac{2\pi}{a^2} \int_0^\infty P_{\text{beam}}(r) r \, dr, \tag{S30}$$

where $a$ is the size of a single emitter. It can be simplified to the form:

$$\langle N_{\text{exc}} \rangle = \frac{2\pi N_e}{a^2 \sigma^2} \int_0^\infty e^{-\frac{r_0^2}{2\sigma^2}} \cdot \sin^2|g(r_0)| r_0 dr_0 \int_0^\infty e^{-\frac{r^2}{2\sigma^2}} I_0\left(\frac{rr_0}{\sigma^2}\right) r \, dr.$$

After the integration, we get:

$$\langle N_{\text{exc}} \rangle = \frac{2\pi N_e}{a^2} \int_0^\infty \sin^2|g(r_0)| r_0 dr_0. \tag{S31}$$

The total number of excited emitters $\langle N_{\text{exc}} \rangle$ as the function of beam size $\sigma$ is shown in Fig. S16. We can see that for beams with $\sigma \geq 10$ nm (which is almost always the case), the averaged number of excited emitters no longer depends on the beam size.

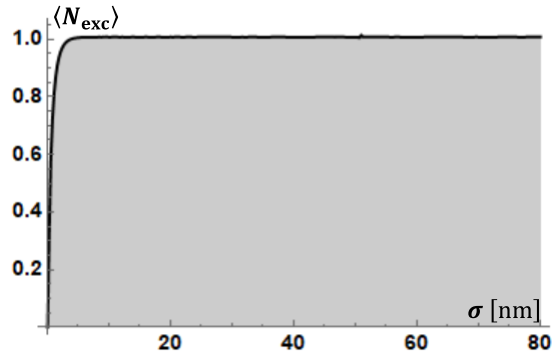

**Fig. S16.** The average number of excitons as the function of the beam size.

However, the number of emitters taking part in the superfluorescence does not equal $\langle N_{\text{exc}} \rangle$. There is a finite distance $R_{\text{max}}$ within which emitters can correlate with each other[18]. Thus, approximately, only the emitters within a circle of radius $R_{\text{max}}$ can simultaneously participate in superfluorescence, as schematically shown in Fig. 17a. Thus, the optimal e-beam size is a constraint both from below and from above.

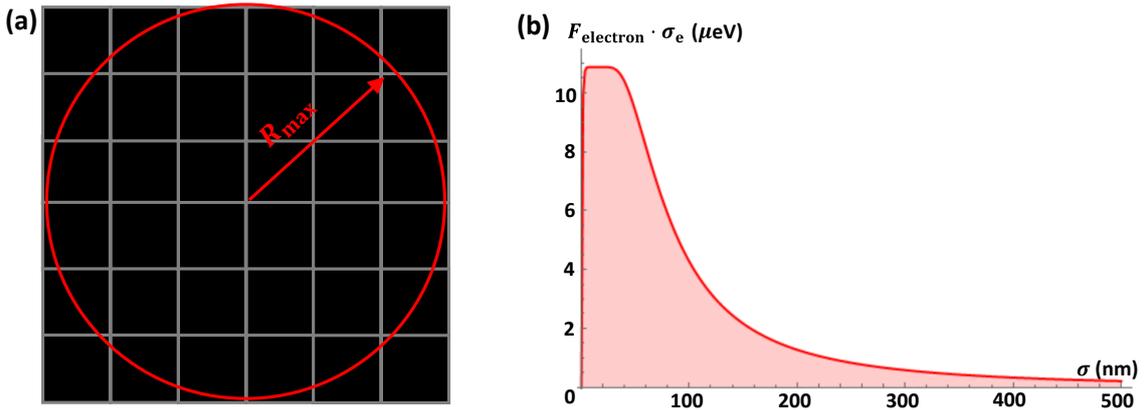

**Fig. S17.** (a) The radius of coherence $R_{\text{max}}$ within which the emitters can build correlations between each other. We estimate that $R_{\text{max}} = 100$ nm in our experiment (for ~20 K). (b) The dependence on beam size of $F_e \cdot \sigma_e$, as shown in Eq. (S33), with interaction constant $g = 0.1$.

For comparison, in the case of light-driven superfluorescence, the number of emitters participating in fluorescence is proportional to the fluence $F_{\text{ph}}$ [in units of $\mu J \cdot m^{-2}$] of the excitation pulse[19], with the time of superfluorescent emission $\tau_{\text{SF}}$ being:

$$\tau_{SF} = \frac{\tau_{SE}}{1+\alpha \cdot \sigma_{ph} \cdot F_{ph}} + \Delta t, \qquad (S32)$$

where $\tau_{SE}$ is the time of emission from a single emitter (conventional spontaneous emission), $\alpha$ is a constant that depends on material properties and not on the type of excitation, $\sigma_{ph}$ is the cross-section parameter of the interaction between light and emitters, and $\Delta t$ is a delay connected with the measurement devices.

In the case of free electrons, we can write the same formula but replace the fluence of light $F_{ph}$ by a fluence parameter for the electrons $F_e$, which describes the energy delivered by the electron beam divided by the beam area. We define the cross-section parameter $\sigma_e$ by the efficiency and area of the electron interaction, such that $F_{electorn} \cdot \sigma_e$ (Fig. S17b) gives the energy of excitons that translates to light emission inside the correlated area $\pi R_{max}^2$:

$$F_e \cdot \sigma_e = \frac{2\pi\hbar\omega_0}{\pi R_{max}^2} \int_0^{R_{max}} P_{beam}(r) r dr = \frac{2\hbar\omega_0 N_e}{\sigma^2 R_{max}^2} \int_0^\infty e^{-\frac{r_0^2}{2\sigma^2}} \sin^2|g(r_0)| r_0 dr_0 \int_0^{R_{max}} e^{-\frac{r^2}{2\sigma^2}} I_0\left(\frac{rr_0}{\sigma^2}\right) r dr. \qquad (S33)$$

Consequently, the time of superfluorescence $\tau_{SF}$ triggered by free electrons can be estimated similarly to light superfluorescence defined in[19]:

$$\tau_{SF} = \frac{\tau_{SE}}{1+\alpha \cdot \sigma_e \cdot F_e} + \Delta t. \qquad (S34)$$

The results of Eq. (S34) are depicted in Fig. S18 and in Fig. 3a in the main text.

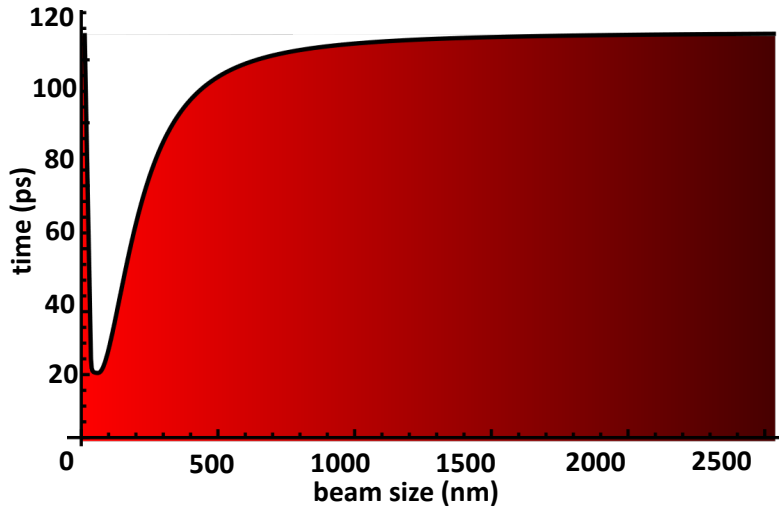

**Fig. S18.** Emission time $\tau_{SF}$ as the function of electron beam size, according to Eq. (S34). To match with the experiment, we take $\tau_{SE} = 128$ ps, and $\Delta t = 1$ ps.

The important prediction of Eq. (S34) is the limit of beam diameter for SF. When the beam is smaller than a single QD, it will not excite more than the QD that it is focused on, and no coherence will be built between individual QDs. In the case of a very large diameter of the beam, we will also not see coherence between the QDs since, most likely, the electrons will be too far from each other. However, in the case when the diameter of the beam is larger than a single quantum dot and smaller than the emitted wavelength, we know that we will see the effect of superfluorescence if there are enough electrons per pulse to excite multiple QDs. This property can be used to investigate the effects of superfluorescence as a function of the number of emitters, by continuously changing the size of the beam to control the number of emitters involved in the process.

Let us further note that with light excitation, it is often the case that multiple decay times should be averaged over (e.g.,[20]). Then, an exponential decay does not fit the data (a Gaussian decay is observed instead). Our data fits to a clear exponential decay, which is a strong indication of the dominance of a single $\tau_{SF}$ value. Physically, this means that the correlations force different emitters to decay simultaneously even when they experience different interaction coupling strengths. This makes sense for superfluorescence (rather than superradiance) because the excitation is highly incoherent, i.e., the phase is lost, and the correlations are all built only during the emission after the excitation strength is no longer important. Altogether, this implies that our experiment observes a unique phenomenon in which the emitters become indistinguishable during their emission, despite being distinguished on excitation.